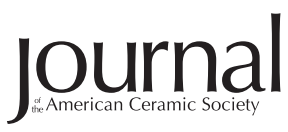

RESEARCH ARTICLE

# Radiation-induced instability of organic-inorganic halide perovskite single crystals

**Ruitian Chen**[1] | **Mingyu Xie**[1] | **Tianyi Lyu**[1] | **Jincong Pang**[2] | **Lewei Zeng**[1] | **Jiahui Zhang**[1] | **Changjun Cheng**[1] | **Renfei Feng**[3] | **Guangda Niu**[2] | **Jiang Tang**[2] | **Yu Zou**[1]

[1]Department of Materials Science and Engineering, University of Toronto, Toronto, Ontario, Canada

[2]Wuhan National Laboratory for Optoelectronics and School of Optical and Electronic Information, Huazhong University of Science and Technology, Wuhan, China

[3]Canadian Light Source, Saskatoon, Saskatchewan, Canada

**Correspondence**
Yu Zou, Department of Materials Science and Engineering, University of Toronto, Toronto, ON M5S 3E4, Canada.
Email: mse.zou@utoronto.ca

**Funding information**
Natural Sciences and Engineering Research Council of Canada, Grant/Award Number: RGPIN-2018-05731; Ontario Early Researcher Award; Canada Foundation for Innovation; CLSI Student Travel Support Program

**Abstract**

Organic-inorganic halide perovskites (OIHPs) are promising optoelectronic materials, but their instability in radiation environments restricts their durability and practical applications. Here, electron and synchrotron X-ray beams are employed with energies in the tens of keV range, individually, to investigate the radiation-induced instability of two types of OIHP single crystals ($FAPbBr_3$ and $MAPbBr_3$). Under the electron beam, the $FAPbBr_3$ sample exhibits 3-point star-style cracks spanning a few micrometers on its surface, and the $MAPbBr_3$ sample shows bricklayer-style cracks extending over tens of micrometers on its surface. Under the X-ray beam, a new composition, $PbBr_2$, is identified in both single crystals. Such cracking phenomena and composition evolutions are attributed to the volatilization of organic components according to energy dispersive X-ray spectroscopy and XRD results. Nanoindentation measurements reveal that beam radiation reduces the Young's modulus of $FAPbBr_3$ (2%) and $MAPbBr_3$ (16%) and increases the hardness of both crystals over 10%. A volume-strain-based mechanism is proposed in which the energy conversion results from the organic cation loss. The difference between volume strain energy and crack formation energy leads to the variation in crack patterns. This study provides insights into the structural and mechanical instability of OIHP single crystals in high-energy beam radiation environments.

**KEYWORDS**
cracking, decomposition, electron beam radiation, perovskites, X-ray synchrotron radiation

## 1 | INTRODUCTION

Halide perovskites, a class of popular optoelectronic and semiconductor materials, are widely employed in photovoltaics, light-emitting diodes, X-ray detectors, and flexible electronics.[1–5] As a subclass of the halide perovskite family, organic-inorganic halide perovskites (OIHPs) are represented by a general chemical formula $ABX_3$, where A is the organic group cation, B is the inorganic cation, and X is the halide anion.[6,7] OIHPs exhibit structural flexibility and tunable functionality that cannot be achieved in all-inorganic halide perovskites.[8,9] The organic components in OIHPs, however, may suffer from low intrinsic and thermodynamic stabilities.[8,10] For example, experimental calorimetric measurements on $MAPbBr_3$ ($MA^+ = CH_3NH_3^+$) and $MAPbI_3$ reveal that they are thermodynamically unstable.[11] The instability and corresponding short lifetime of OIHPs are obstacles that impede their commercial viability for long-period operations.[12,13] Therefore, the evaluation and characterization of the instability of OIHPs under various external stimuli is essential to understand the underlying mechanisms and pave the way for enhancing their stability for long-duration applications.

The instability of OIHPs under high-energy radiation has been widely reported. For example, grain-boundary cracks with “mud-cracking” patterns appeared on the surfaces of OIHP polycrystalline thin films during scanning electron microscopy (SEM) imaging.[14,15] Structural damage and degradation were also observed in OIHP thin films exposed to electron beam (e-beam) and X-ray radiation.[16–20] Moreover, proton radiation could induce lattice displacement and defect formation,[21–23] while alpha particles and gamma rays could trigger phase transitions and degrade the power conversion efficiency of OIHP polycrystalline films.[24–27] Compared to polycrystalline thin films, OIHP single crystals are free of grain boundaries and exhibit lower defect density, higher carrier mobility, longer carrier lifetimes, and larger absorption coefficients.[28–30] Therefore, the influence of radiation on the surface of OIHP single crystals is also worthy of attention. The e-beam effect on OIHPs is noteworthy, as their morphology, orientation, and crystal growth process are based on electron microscopy characterization.[31–34] Recent studies have reported that the e-beam could induce defects and decomposition in $MAPbBr_3$ single crystals because of organic species volatilization.[35,36] The morphology damage in $MAPbI_3$ single crystals was observed under e-beam radiation using transmission electron microscopy (TEM).[37] Besides, the instability and decomposition of single-crystalline $MAPbBr_3$ and $MAPbI_3$ were detected using high-resolution TEM (HRTEM) and scanning transmission electron microscopy (STEM) at the atomic level.[38–40] In addition, the stability and performance of OIHP single crystals under the X-ray beam radiation are also critical issues as they are effective candidates for photo-imaging and X-ray detectors.[3,41] For instance, the X-ray beam-induced degradation of $MAPbBr_3$ single crystals was inevitable when the samples were tested under X-ray radiation with monitoring by X-ray photoelectron spectroscopy (XPS).[42,43] The decomposition of $FAPbBr_3$ ($FA^+ = CH(NH_2)_2^+$) and $MAPbBr_3$ single crystals were also detected under synchrotron soft X-ray.[44,45] Although many previous studies have investigated the instability issues of OIHP single crystals under e-beam and X-ray beam radiation, further in-situ exploration is needed to capture the real-time evolution of these processes and to understand the disparities among crystals with varying compositions, as well as the distinctions and interconnections between different types of beam radiation effects. Additionally, the impact of radiation on mechanical performance is crucial, as the mechanics-coupled stability of OIHPs plays a vital role in their durability and commercial viability.[12,46]

In our study, we perform in-situ observations of both e-beam-induced cracking in SEM and synchrotron X-ray beam-induced composition evolutions of two different OIHP single crystals, including $FAPbBr_3$ and $MAPbBr_3$. We reveal the overall unstable process of OIHP single crystals under radiation and directly detect the volatilization of organic species experimentally. Besides, we also test the Young’s modulus ($E$) and hardness ($H$) on the as-grown and radiated areas of both crystals, using the nanoindentation method, to investigate the impact of e-beam-induced cracking on their mechanical performance. Based on the loss of organic components, we unveil a volume-strain-based mechanism in OIHP single crystals under beam radiation from the perspective of energy conversion.

## 2 | METHODS

### 2.1 | Synthesis of halide perovskite single crystals

For the preparation of $MAPbBr_3$ single crystals, the precursor solution was prepared by mixing 311.3 mg of $CH_3NH_3Br$ (0.65 mmol) and 238.6 mg of $PbBr_2$ (0.65 mmol) in 0.6 mL of Dimethylformamide solvent. The solution was gradually heated to 90°C at a rate of 5°C and maintained at this temperature for 6 h, resulting in the formation of seed crystals. To increase the crystal size, the temperature was further raised to 110°C at a rate of 10°C and held steady overnight. This process yielded $MAPbBr_3$ single crystals with dimensions of up to 2 mm × 2 mm × 3 mm. The crystals were then removed from the solution and stored at room temperature. $FAPbBr_3$ single crystals were grown from a precursor solution prepared by dissolving FABr and $PbBr_2$ in $\gamma$-butyrolactone, stirred, and filtered. Seed-assisted slow temperature elevation from 40°C to 50°C at 0.1°C $h^1$ enabled crystal growth over several days. $CsPbBr_3$ single crystals were synthesized from stoichiometric CsBr and $PbBr_2$ purified by zone melting, followed by complete reaction in a swing furnace and single crystal growth via the Bridgman method. The ingots were then cooled and processed through cutting, grinding, and polishing to obtain high-quality crystal samples. Detailed parameters for the synthesis of $FAPbBr_3$ and $CsPbBr_3$ single crystals are as reported in our previous study.[47] Experiments were conducted on the (100) plane of $FAPbBr_3$ and $MAPbBr_3$ single crystals, and (101) plane of the $CsPbBr_3$ single crystal. X-ray diffraction (XRD) results are shown in Figure S1.

**TABLE 1** Radiation conditions on $FAPbBr_3$ and $MAPbBr_3$ single crystals.

| Beam type | Beam energy | Beam density[a] |
|---|---|---|
| e-beam | 10 keV | 7E6 e/μm$^3$; 2E7 e/μm$^3$; 6E7 e/μm$^3$ |
| | 12 keV | 6E7 e/μm$^3$ |
| | 15 keV | 6E7 e/μm$^3$ |
| X-ray beam | 11 keV | 5E9 ph/μm$^3$ |
| | 14 keV | 5E9 ph/μm$^3$ |
| | 17 keV | 5E9 ph/μm$^3$ |

[a] e, elementary electric charge; ph, photon number.

## 2.2 | In-situ beam radiation

In-situ SEM imaging and energy dispersive X-ray spectroscopy (EDS) analysis were performed using the Hitachi FE-SEM SU 7000 in the low vacuum mode with a pressure of 5 Pa. In-situ SEM imaging was conducted under different accelerated voltages, including 10, 12, and 15 kV, and different beam densities by tuning the magnifications. The accelerated voltage for EDS analysis was 15 kV. In-situ grazing incidence X-ray diffraction (GI-XRD) was performed in the Very Sensitive Elemental and Structural Probe Employing Radiation from a Synchrotron (VESPERS) beamline at the Canadian Light Source, and XRD patterns were analyzed based on ALBULA and XMAS software.[48,49] The e-beam and X-ray beam radiation energies and beam densities are summarized in Table 1. The detailed calculation methods for beam density are discussed in Supplementary Materials, and the penetration depths of beam radiation with different energies are listed in Table S1.

## 2.3 | Nanoindentation

Nanoindentation tests were performed in the KLA iMicro Nanoindentation system equipped with a Berkovich diamond indenter tip (Synton-MDP) at room temperature.[47] The tip shape function was calibrated by testing the fused silica standard. The drift rates were controlled below 0.1 nm/s, and the sampling frequency was 110 Hz. Young's modulus and hardness were measured by the Oliver and Pharr method in the continuous stiffness measurement mode with an indentation depth of 1000 nm.[50,51] Multiple tests were repeated to ensure reliable results. Figure 1 shows the overall scheme for the experiment procedures.

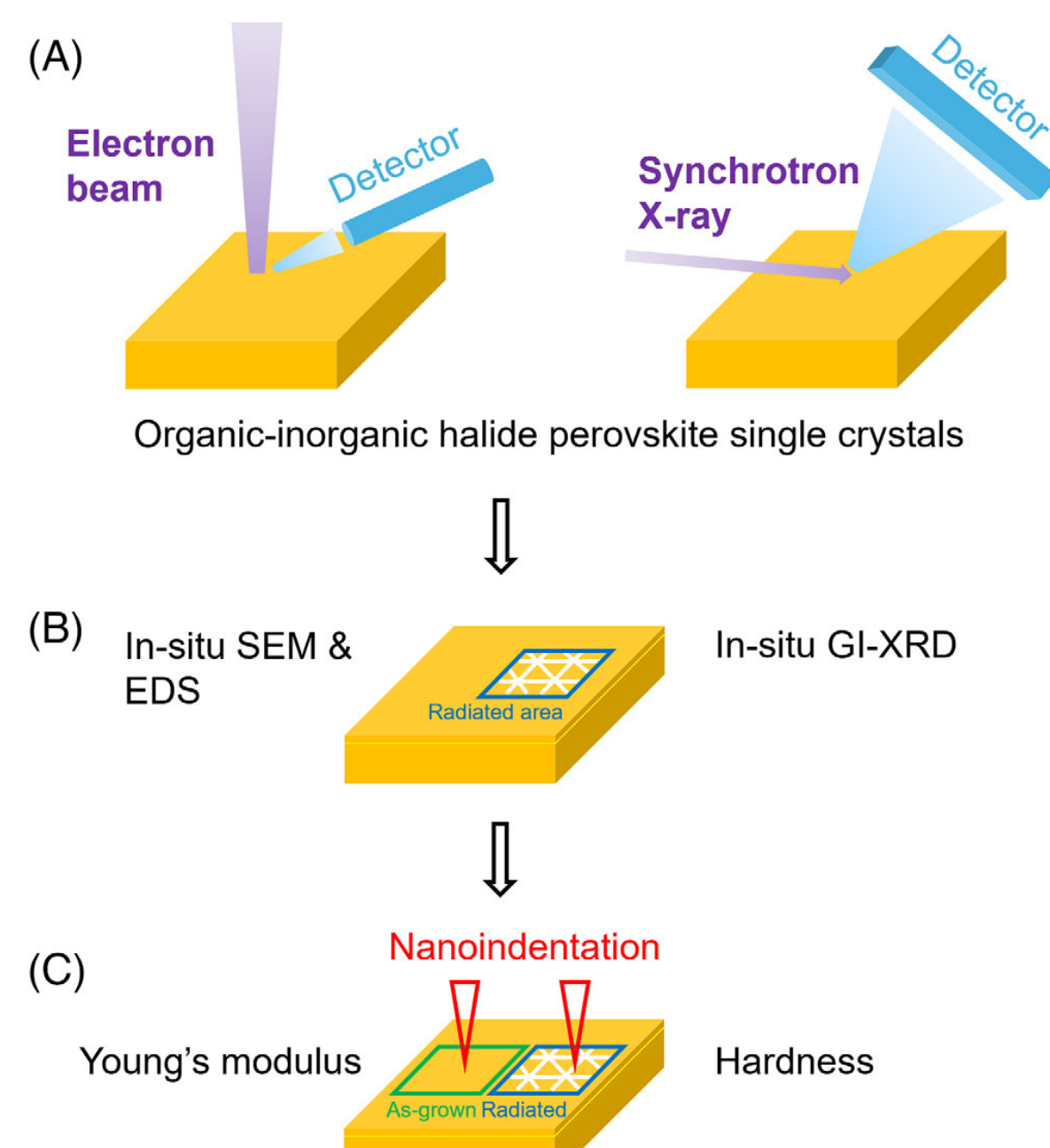

**FIGURE 1** Schematic illustration of the experimental procedures for (A) electron and X-ray beam radiation, (B) in-situ characterization, and (C) nanoindentation tests on the surface of OIHP single crystals.

# 3 | RESULTS

## 3.1 | In-situ SEM imaging and EDS analysis

Figure 2 shows the crack initiation and propagation on the surface of $FAPbBr_3$ and $MAPbBr_3$ single crystals under e-beam radiation of 2E7 e/μm$^3$ @ 10 keV as the increase of radiation time (see Videos S1 and S2 for SEM imaging of $FAPbBr_3$ and $MAPbBr_3$ under different e-beam energies and densities). The crack patterns are different between these two types of single crystals. Analogous to mud desiccation crack patterns, there are 3-point star-style cracks on the $FAPbBr_3$ crystal surface and bricklayer-style cracks on the $MAPbBr_3$ crystal surface.[52,53] For $FAPbBr_3$, the longer the radiation exposure, the more cracks form, but their lengths are almost unchanged over time. In contrast, cracks grow longer quickly in $MAPbBr_3$, but the number of cracks is fewer than those in $FAPbBr_3$. These phenomena indicate that e-beam-induced cracking occurs on the surface of OIHP single crystals without grain boundaries, and the crack patterns are distinct in OIHPs with different organic cations.

Figure 3 shows the percentage of crack area (i.e., the percentage of crack occupation area in the total scanned area) and crack length changes over radiation time under various circumstances. Crack initiation and propagation under e-beam radiation can be categorized into three stages. In Stage I, OIHP single crystals remain stable, and no cracks appear. Both crystals keep in this stage for durations ranging from several tens to approximately 200 s, but only under low e-beam density and energy (7E6 e/μm$^3$ @ 10 keV and 2E7 e/μm$^3$ @ 10 keV). In Stage II, the cracks initiate and propagate, and the percentage of crack area rises rapidly. In Stage III, the increase in the percentage of crack area slows down. In Stages II and III, the crack length quickly reaches its upper limit, with both the median and longest cracks showing only slight fluctuations as more cracks form. The sudden increase followed by a reduction in the median crack length of $MAPbBr_3$ under low e-beam density and energy is attributed to the initial emergence of several long cracks, followed by the formation of shorter cracks that intersect with the earlier ones.

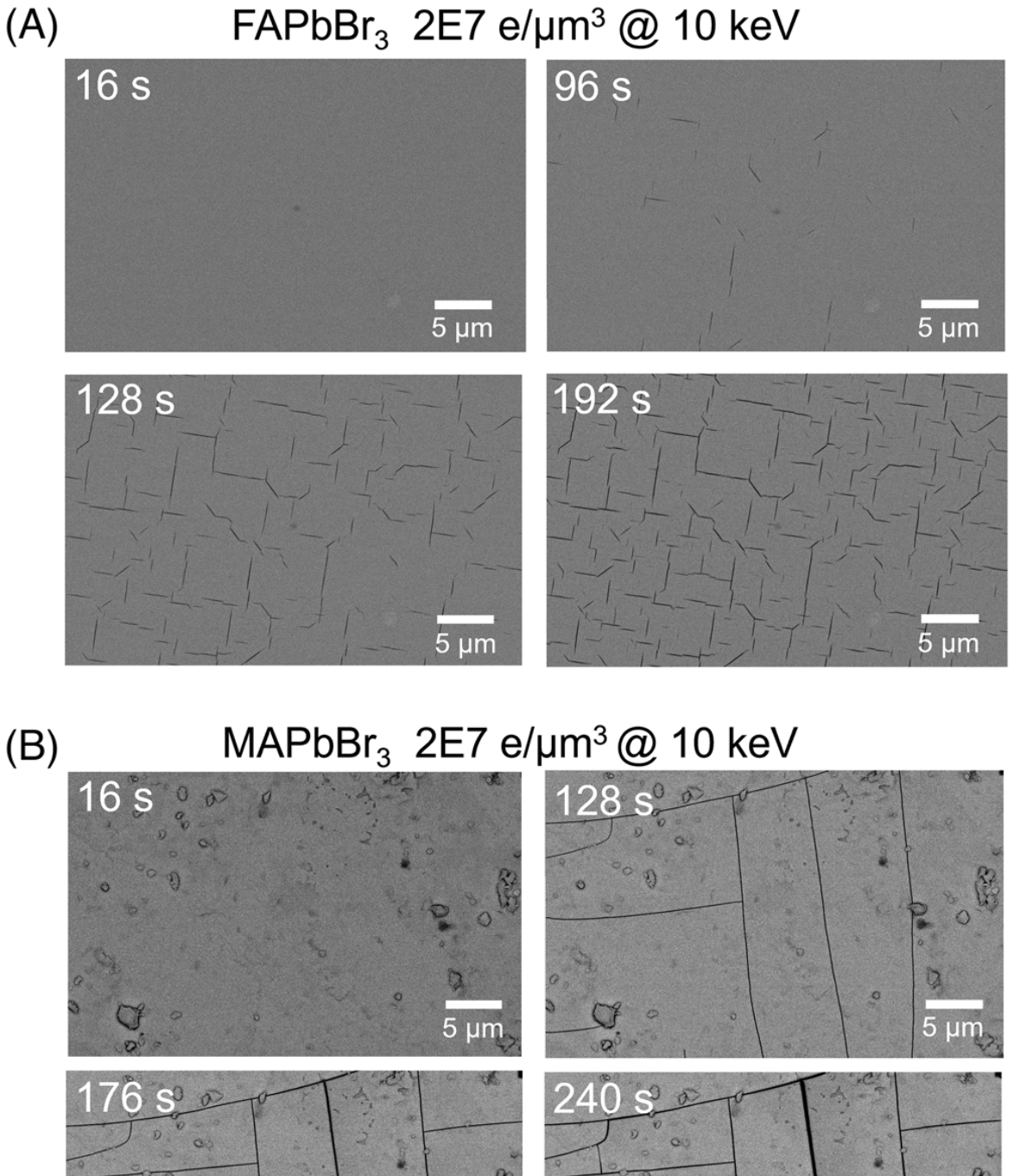

**FIGURE 2** SEM images of (A) $FAPbBr_3$ and (B) $MAPbBr_3$ single crystals under the e-beam radiation at 2E7 e/μm$^3$ @ 10 keV.

There are some common features in both crystals: for example, as the beam density increases, the duration of Stage I decreases. Crack growth occurs at the beginning when the beam density reaches 6E7 e/μm$^3$. Besides, the higher the beam density or energy, the faster the percentage of crack area increases. However, the two crystals also exhibit distinct trends in the percentage of crack area and crack length, along with their distinct crack patterns. The beam density influences more significantly on the percentage of crack area of $FAPbBr_3$, whereas the beam energy affects more significantly on the percentage of crack area of $MAPbBr_3$. The beam density indicates how focused the e-beam scanning area is, while the beam energy affects the electron's incident depth. The differing crack patterns in the two crystals are each more sensitive to one of the two factors, suggesting that the cracking processes in these two crystals are not identical. In addition, Figure 4 shows the crack morphology of both single crystals in Stage III after e-beam radiation under different e-beam energies and densities. The percentage of crack area of $MAPbBr_3$ increases more than that of $FAPbBr_3$ in Stage III because there is an apparent widening of cracks on the $MAPbBr_3$ crystal surface. Because of the much longer and wider cracks, $MAPbBr_3$ obtains a higher percentage of crack area than $FAPbBr_3$ at the high e-beam energy of 15 keV. Crack lengths depend on the beam density in both crystals: their maximum values decrease with increasing beam density (i.e., higher scanning magnification and smaller scanning area), but remain similar under the same beam density, respectively. Crack lengths in $FAPbBr_3$ are substantially shorter than those in $MAPbBr_3$ under the same conditions, and the cracks in $MAPbBr_3$ can extend to the boundaries of the scanning area.

To unveil the cracking mechanisms, the all-inorganic halide perovskite $CsPbBr_3$ is characterized for comparison. There is no crack on the surface of the $CsPbBr_3$ single crystal after an e-beam radiation of 240 s under 6E7 e/μm$^3$ @ 10 keV, as shown in Figure S2, which suggests that the unstable organic cations play a key role in crack growth. The results are also consistent with literature studies on halide perovskite polycrystalline films, which show that $CsPbBr_3$ thin films exhibit higher e-beam stability than $MAPbBr_3$.[54,55] Figure 5 shows the in-situ EDS analysis results and reveals that the contents of C and N elements in both $FAPbBr_3$ and $MAPbBr_3$ single crystals decrease under 150-s e-beam radiation. Cracks after EDS experiments are observed in Figure S3. Typically, contamination from residual impurities in the SEM chamber leads to a gradual increase in the EDS signals of C contents over time,[56,57] but the remarkable decrease of C and N contents indicates that the organic components in OIHP single crystals volatilize during e-beam radiation. Besides, the amount of C and N elements in $MAPbBr_3$ reduces more and faster than that in $FAPbBr_3$, corresponding to the higher percentage of crack area of $MAPbBr_3$. The contents of C and N in $MAPbBr_3$ tend to remain unchanged after 50 s, while they continue decreasing in $FAPbBr_3$ after 100 s. The element content fluctuations also increase after long-duration radiation due to contamination.[57]

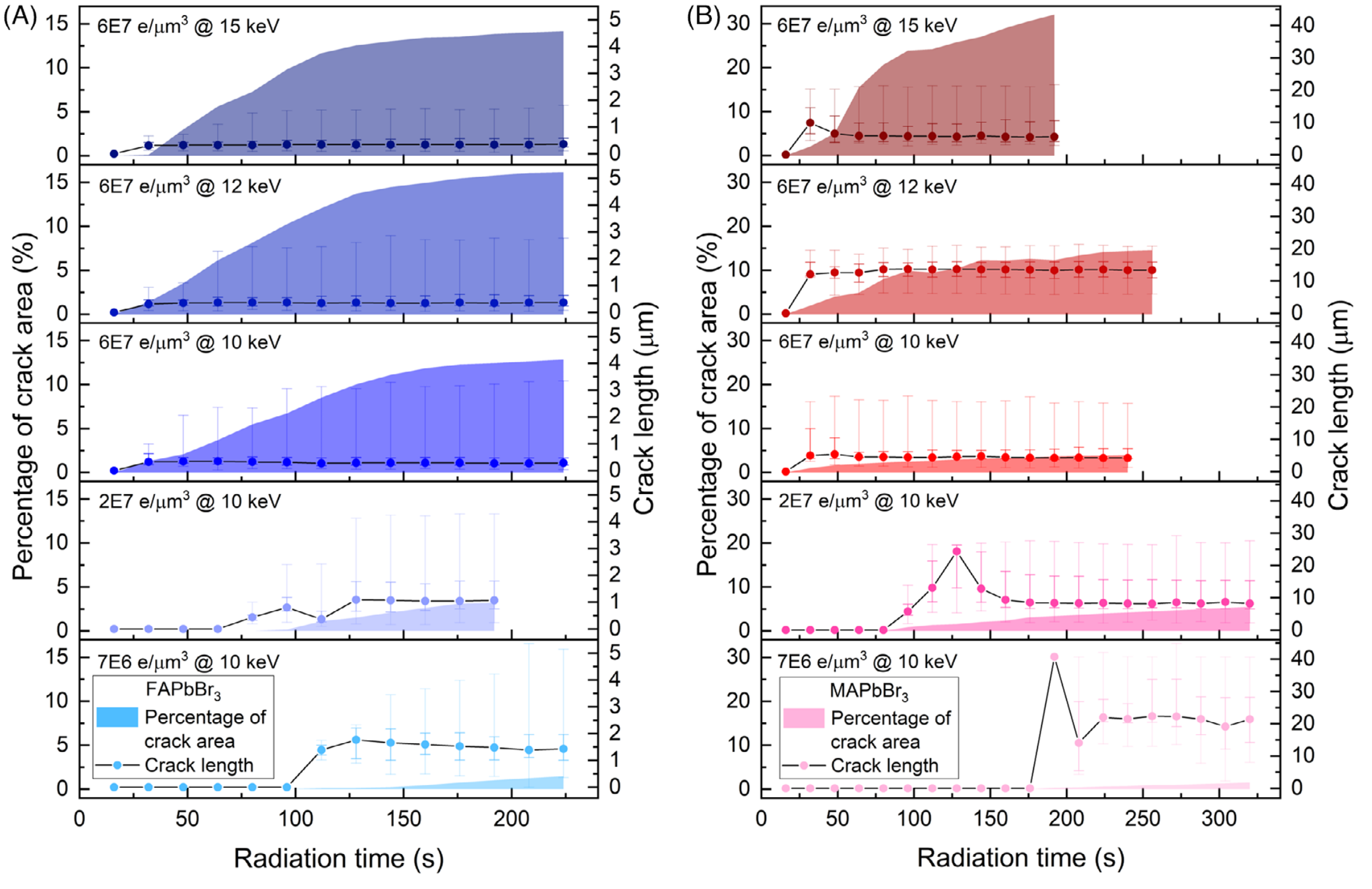

**FIGURE 3** The percentage of crack area and crack length over the radiation time of (A) $FAPbBr_3$ and (B) $MAPbBr_3$ single crystals under various e-beam densities and energies. For crack length, the figure shows the median (points), interquartile range (inner bars: first and third quartiles), and full range (outer bars: minimum to maximum).

## 3.2 | In-situ synchrotron XRD analysis

The synchrotron X-ray beam provides high-energy radiation in a small area, and the synchrotron GI-XRD is an ideal tool to analyze the composition evolution process on the surface of OIHP single crystals.[48,58] Figure 6 shows the in-situ GI-XRD results of the $FAPbBr_3$ and $MAPbBr_3$ single crystals under 5E9 ph/μm$^3$ @ 11 and 17 keV. Results for X-ray beam energy at 14 keV of both crystals are shown in Figure S4, and the XRD pattern evolutions are displayed in Videos S3 and S4. As radiation time increases, new phases appear in both crystals, corresponding to the crystal structure of $PbBr_2$. In comparison, no new peaks are observed in the GI-XRD results of the $CsPbBr_3$ single crystal under X-ray beam radiation of 5E9 ph/μm$^3$ @ 11 keV over 1080 s, as shown in Figure S5. The high phase stability of $CsPbBr_3$ under an X-ray beam is consistent with the literature.[59] In contrast, the generation of the new composition without organic parts and the weakening of peak intensities for the original structure in OIHP single crystals indicate that, in addition to the e-beam, the high-energy X-ray beam can also induce the volatilization of organic components and lead to their structural instability.

Figure 7 presents the percentage of peak intensity ratios to show the composition and structural evolution process of $FAPbBr_3$ and $MAPbBr_3$ single crystals under X-ray beam radiation. The peak intensity ratios between some $PbBr_2$ peaks and $FAPbBr_3$ or $MAPbBr_3$ (100) peak over radiation time are calculated, and the percentage is normalized by dividing by the last ratio value (detailed calculation methods are given in Supplementary Materials). Besides, Figure S6 presents the intensity ratios between $PbBr_2$ (111) peak and $FAPbBr_3$ or $MAPbBr_3$ (100) peak under different X-ray beam energies. Comparable to the percentage of crack area changes under e-beam radiation, the changes of X-ray beam-induced peak intensity ratios can also be categorized into three stages. In Stage I, OIHP single crystals remain stable, and there are no new phases or peaks. At the low X-ray beam energy of 11 keV, Stage I lasts for several hundreds of seconds. However, at 14 and 17 keV, Stage I does not last long for both crystals. In Stage II, $PbBr_2$ peaks appear, and the peak intensity ratio increases quickly. In Stage III, the rate of increase in intensity ratios slows, and the composition evolution approaches saturation when the X-ray beam energy is high. In general, the new $PbBr_2$ peaks emerge more rapidly under higher beam energy, and the intensity reaches saturation more quickly. Composition evolutions in $MAPbBr_3$ are much more intensive than those in $FAPbBr_3$ induced by X-ray beam, as evidenced by the higher intensity ratios shown in Figure S6, which is consistent with the variation in the percentage of crack area under the e-beam radiation.

**FIGURE 4** SEM images of (A) $FAPbBr_3$ and (B) $MAPbBr_3$ in Stage III showing crack morphology after e-beam radiation under different e-beam energies and densities.

### 3.3 | Nanoindentation studies

Figure 8A,B shows the SEM images of $FAPbBr_3$ and $MAPbBr_3$ single crystals within and beyond the radiation area after an e-beam exposure of 180 s, indicating that cracks only appear in the region exposed to the e-beam. To investigate the influence of e-beam-induced cracking on the mechanical performance of OIHP single crystals, nanoindentation tests are conducted to measure $E$ and $H$ on both the as-grown surface and the radiated surface areas of the two crystals.

As shown in Figure 8C,D, the radiated area exhibits lower $E$ and higher $H$ values compared to the as-grown area in both crystals. Since the indent size encompasses numerous cracks, the decreased $E$ values could be attributed to reduced stiffness and strength in the radiated area, and the accumulation of cracks disrupts the continuity of the crystal lattice. In contrast, the increased $H$ values might be due to the loss of organic components and changes of the bonding types, which reduces the flexibility of the crystal structure.[60–62] In addition, the changes of $E$ and $H$ values are more pronounced in $MAPbBr_3$ (16% decrease in $E$ and 17% increase in $H$) compared to $FAPbBr_3$ (2% decrease in $E$ and 11% increase in $H$). This discrepancy arises from the more extensive crack formation on the surface of $MAPbBr_3$, where the cracks are significantly longer, wider, and more irregular, even leading to surface wrapping and increased roughness. The mechanical integrity of OIHP single crystals is compromised under radiation exposure, which may impact their long-term reliability in practical applications.

## 4 | DISCUSSION

### 4.1 | Mechanisms of cracking and composition evolutions

The OIHP single crystals experience crack initiation and propagation under e-beam radiation, and composition evolutions under X-ray beam radiation. Both phenomena are associated with the volatilization of organic components. Figure 9A shows the schematic process from the macroscopic view, resembling the desiccation cracking of clay soil during drying, and Figure 9B illustrates the corresponding structural and composition evolutions at the lattice scale from the perspective of energy conversion. In clay soil, volume shrinkage occurs as water evaporates. When this shrinkage is constrained, tensile stress develops within the soil. Once the tensile stress exceeds the soil's strength, cracks initiate.[63–65] For $FAPbBr_3$ and $MAPbBr_3$ single crystals, the high-energy beam in the tens of keV range is able to break chemical bonds and induce the volatilization of $FA^+$ and $MA^+$ organic cations on the crystal surface, allowing light elements such as C, N, and H to escape.[34] The loss of organic cations generates numerous vacancies and a negative volume strain, leading to structural instability. The unstable intermediate structure then transforms into products with stable structures, while the energy released from the volume strain is converted into crack formation energy. Lead dihalides have been identified as the degradation product of OIHPs under e-beam radiation.[38,39,66] Our GI-XRD results also confirm the formation of $PbBr_2$ under X-ray beam radiation, consistent with previous reports.[43] Although $PbBr_2$ may further decompose to metallic Pb under X-ray, Pb would readily recombine into compounds in the ambient environment.[42] Therefore, we propose $PbBr_2$ as the primary decomposition product of $FAPbBr_3$ and $MAPbBr_3$ single crystals.

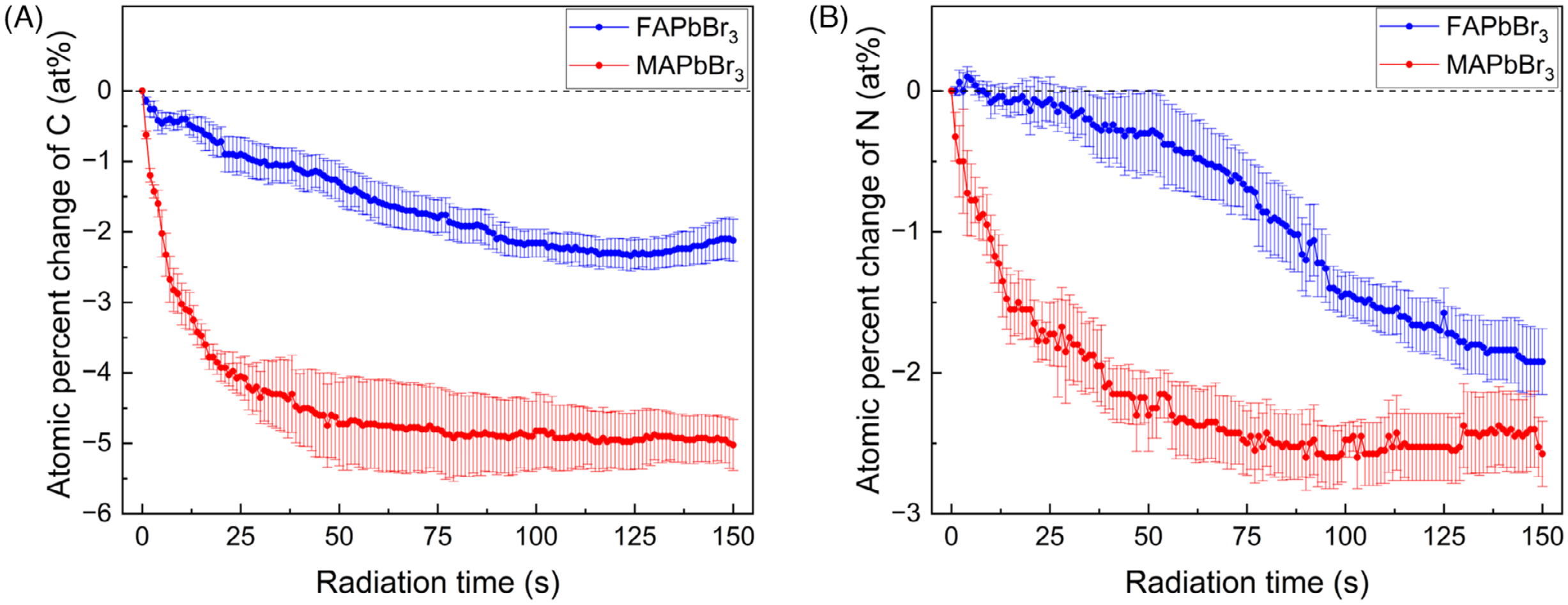

**FIGURE 5** In-situ EDS analysis of the atomic percent (at%) changes of (A) C and (B) N elements in $FAPbBr_3$ and $MAPbBr_3$ single crystals under the e-beam radiation up to 150 s (the starting point at 0 s is set as the reference).

The volume strain energy can be described as follows:

$$\Delta\ E_1 = \frac{1}{2}\,B\left(\rho\frac{\Delta V}{V}\right)^2, \tag{1}$$

where $B$ is the bulk modulus, $\rho$ is the percentage of crack area, and $\Delta V/V$ is the relative volume change.

The crack formation energy can be described as follows, according to the Griffith theory of fracture[67]:

$$\Delta\ E_2 = 4a\gamma n, \tag{2}$$

where $2a$ is the crack length, $\gamma$ is the surface tension ($\sim$ 1 N/m[66,68]), and $n$ is the number of cracks per unit area. Some structural parameters are listed in Table S2.

Based on the $\rho$, $2a$, and $n$ values obtained from SEM imaging, $\Delta\,E_1$ and $\Delta\,E_2$ can be estimated for both single crystals. For $FAPbBr_3$, the volume strain energy and crack formation energy are at the same order of magnitude, so the strain energy is primarily dissipated through crack propagation. In contrast, for $MAPbBr_3$, the crack formation energy is one to two orders of magnitude lower than the volume strain energy. Because the crack lengths in $MAPbBr_3$ have approached the limit of the scanning region, the excess volume strain energy would contribute not only to crack propagation but also to notable crack widening, as shown in Figure 4. In addition, crack opening and surface wrapping is significant in $MAPbBr_3$ under high e-beam energy and density, as high e-beam density introduces more localized deformation, and high e-beam energy enhances volatilization of organic cations from deep regions. If the SEM magnification is sufficiently high, reducing the scanning region size such that cracks in $FAPbBr_3$ can reach the boundaries of the scanning area, the brick-layer pattern may also be displayed on the surface of the $FAPbBr_3$ single crystal.

The differences in intrinsic mechanical properties between $FAPbBr_3$ and $MAPbBr_3$ single crystals would influence their crack growth behavior. $FAPbBr_3$ exhibits a lower Young's modulus and hardness than $MAPbBr_3$, as shown in Figure 8. Therefore, there is relatively reduced resistance to localized surface damage in $FAPbBr_3$, leading to the simultaneous initiation of multiple cracks that propagate radially.[69] Besides, Figure S7 indicates that $FAPbBr_3$ also possesses lower fracture toughness than $MAPbBr_3$. The lower fracture toughness implies a smaller process zone, resulting in limited propagation and rapid termination of cracks in $FAPbBr_3$.[70] Overall, the lower threshold for crack initiation due to lower Yong's modulus and hardness of $FAPbBr_3$ leads to concurrent activation of 3-point star-style cracks, and its higher brittleness causes shorter crack lengths. In contrast, the small number of bricklayer-style cracks in $MAPbBr_3$ suggests greater resistance to crack initiation but an enhanced capacity for crack growth once formed, due to more effective stress accommodation at the crack tip, consistent with the higher Young's modulus, hardness, and fracture toughness of $MAPbBr_3$.

The inherent structural and stability differences between $FA^+$ and $MA^+$ organic cations would contribute to their distinct volatilization processes. There are two C-N bonds with different bond strengths in the $FA^+$ cation, but only one C-N bond in the $MA^+$ cation. Therefore, the e-beam energy has a greater impact on the bond breaking and volatilization rate of $MA^+$, while the bond breaking of $FA^+$ is more influenced by the e-beam density. Some atom distances of $FA^+$ and $MA^+$ in their unit cells, which reflect the bond lengths, are listed in Table S3. Because the $MA^+$ cation size is smaller than the $FA^+$ cation, its bond breaking and escape from the lattice is comparatively easier. In addition, it has been reported that $MA^+$ exhibits lower thermal and chemical stability than $FA^+$. $MA^+$ is more acidic and could readily undergo fragmentation and deprotonation under operating conditions.[71–74] Owing to its smaller size and higher reactivity, the volatilization of $MA^+$ is more pronounced than that of $FA^+$, leading to faster cracking and a higher crack area percentage in $MAPbBr_3$ under e-beam radiation. However, under X-ray beam radiation at 11 keV, the duration of the stable period (Stage I) is shorter for $FAPbBr_3$ than for $MAPbBr_3$. Unlike e-beam, which can directly break chemical bonds, the X-ray beam primarily acts as an indirect energy source that excites bonding and lattice vibrations. In $FAPbBr_3$, the larger size of $FA^+$ leads to weaker bond strength within the inorganic framework, along with weaker hydrogen bonding interactions between $FA^+$ and the framework compared to $MA^+$.[46] Therefore, $FAPbBr_3$ is more vulnerable to X-ray-induced degradation. The distinction ways in which electron and X-ray interact with the lattice also explain why Stage I lasts longer for both crystals under X-ray beam radiation than under e-beam radiation at 11 keV.

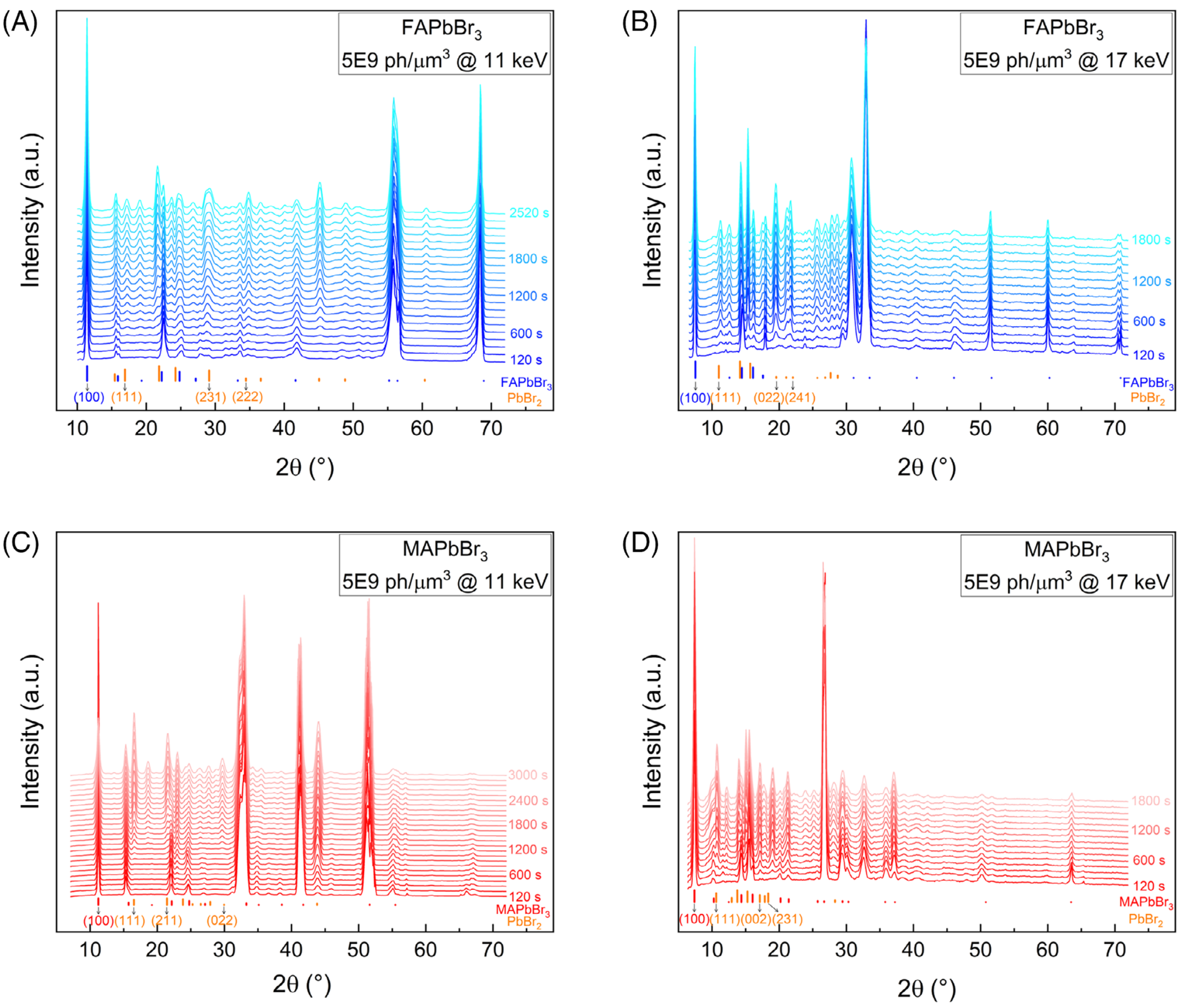

**FIGURE 6** In-situ synchrotron GI-XRD results of (A, B) $FAPbBr_3$ and (C, D) $MAPbBr_3$ single crystals under 5E9 ph/μm$^3$ @ 11 and 17 keV for various durations (typical standard XRD peaks are indicated at the bottom of the figures).

Ferroelastic domains are commonly observed as a mechanism for strain release in halide perovskites when there are phase transitions induced by temperature variations or external stress.[75–78] However, our in-situ SEM imaging and synchrotron GI-XRD results show neither the characteristic stripe-like wall structures in the SEM images nor the peak splitting in the XRD patterns, indicating that fer-roelastic domain formation does not play a major role in the release of strain under beam radiation.[76,79] Nonetheless, we acknowledge that more sensitive local probes such as piezoresponse force microscopy and confocal micro-Raman spectroscopy may reveal subtler ferroelastic rearrangements and represent a promising direction for future investigation.[75,79]

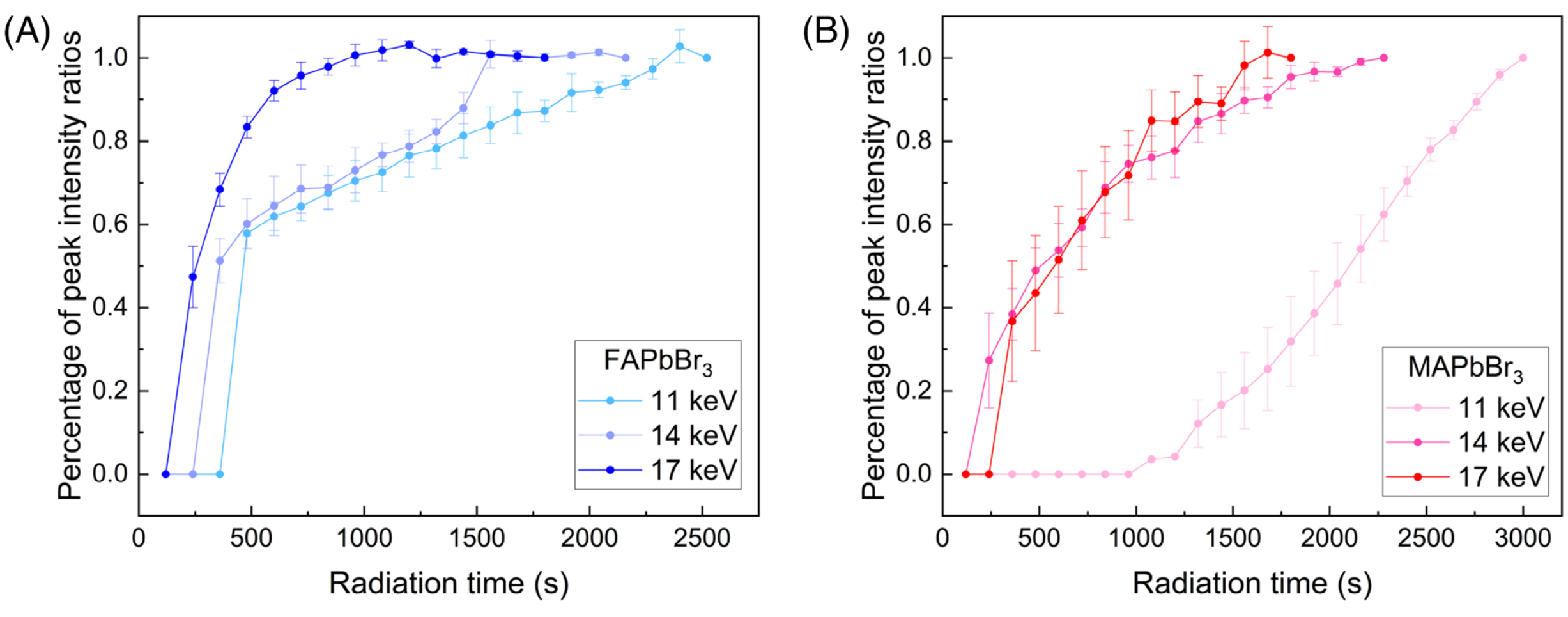

**FIGURE 7** Percentage of peak intensity ratios between some $PbBr_2$ peaks and (A) $FAPbBr_3$ (100) peak or (B) $MAPbBr_3$ (100) peak under different X-ray beam radiation energies ($PbBr_2$ peaks used for calculation are indicated in the figures of GI-XRD results).

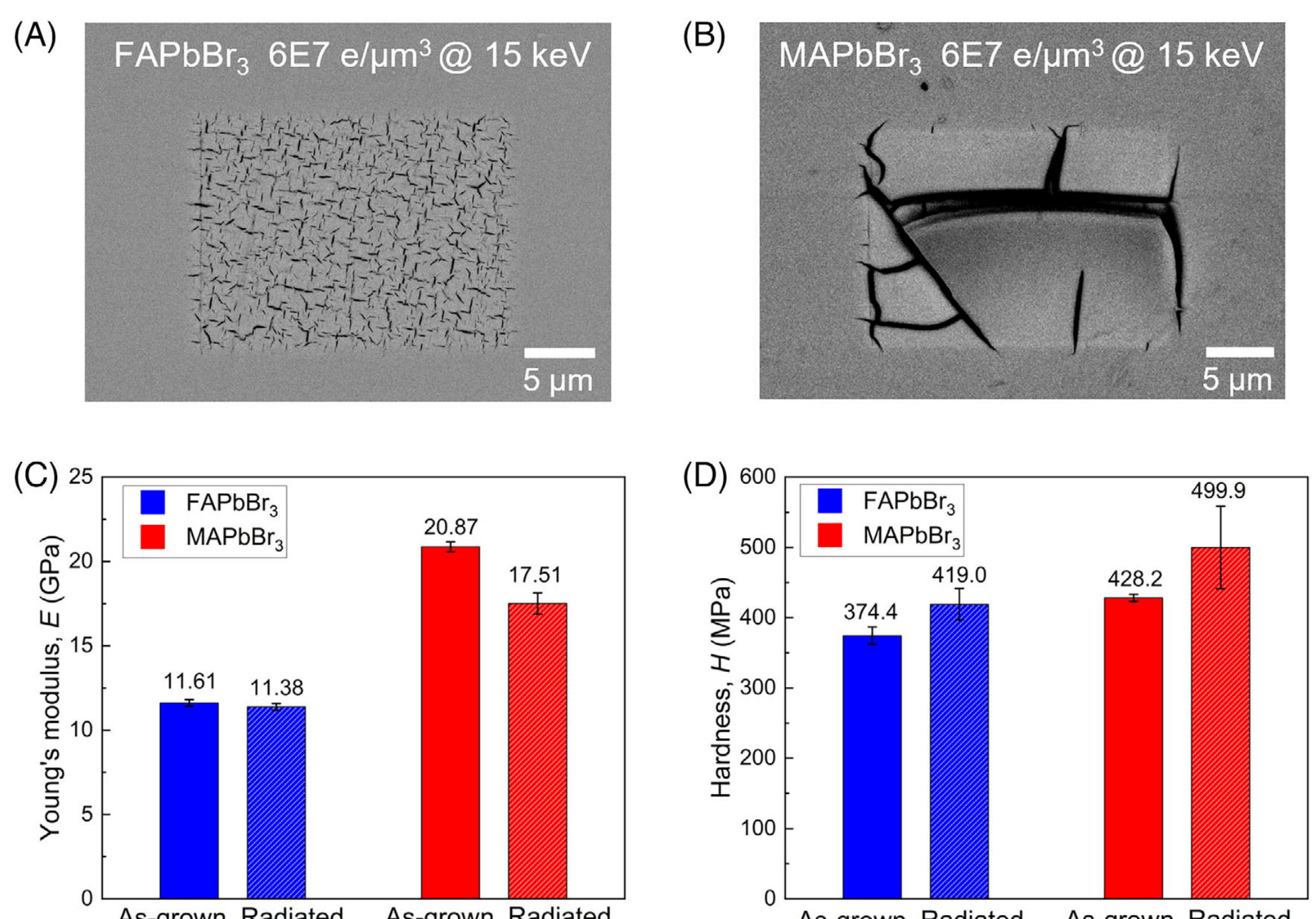

**FIGURE 8** SEM images of (A) $FAPbBr_3$ and (B) $MAPbBr_3$ single crystals within and beyond the radiation area after an e-beam exposure of 180 s. (C) Young's modulus (*E*) and (D) hardness (*H*) of both as-grown and radiated surface areas of $FAPbBr_3$ and $MAPbBr_3$ single crystals.

## 4.2 | Comparisons of radiation-induced OIHP single crystal instability

In our work, we choose similar e-beam and X-ray beam energies in the tens of keV range to enable a meaningful comparison between the two radiation types. Radiation in this energy range is close to common operating conditions in SEM and XRD and minimizes surface-dominated effects with penetration depths of several micrometers (Table S1). In addition, energies in the tens of keV range effectively induce volatilization of organic cations and degradation of OIHP single crystals while still allowing simultaneous characterization using SEM and GI-XRD.

**FIGURE 9** Mechanisms of beam radiation-induced cracking and composition evolutions on the surface of $FAPbBr_3$ and $MAPbBr_3$ single crystals. (A) Schematic process from the macroscopic view. (B) Schematic of lattice structure evolutions.

Table 2 compares the instability of OIHP single crystals under both e-beam and X-ray beam radiation with different energies, incorporating some representative previous studies along with our work. By utilizing appropriate characterization methods under different beam conditions, the instability of OIHP single crystals is well elucidated. For the e-beam radiation, previous studies have revealed the bubble-like flaws in $MAPbI_3$ through the TEM observation[37] and decomposition pathway of $MAPbBr_3$ and $MAPbI_3$ in HRTEM and STEM.[38–40] Our work shows a more direct phenomenon of cracking on the surface of $FAPbBr_3$ and $MAPbBr_3$ single crystals in SEM and provides a more convenient approach to demonstrating the decomposition through EDS analysis. For X-ray beam radiation, recent studies have used XPS to monitor elemental ratios and electrical structure changes of OIHP single crystals under different beam energies.[42–45] In our work, we utilize the synchrotron high-energy X-ray beam to achieve inducing the decomposition of $FAPbBr_3$ and $MAPbBr_3$ single crystals while simultaneously collecting GI-XRD patterns. The observed changes in GI-XRD patterns effectively demonstrate composition evolutions on the OIHP single crystal surface.

The X-ray beam-induced degradation of the $MAPbBr_3$ single crystal under different atmospheres has shown that $N_2$ can provide protection, while the crystal exhibits minimal sensitivity to $O_2$ exposure.[42] Therefore, the synchrotron X-ray beam-induced instability in our work, conducted under ambient conditions, is unlikely to be significantly affected by the environmental atmosphere. Besides, prolonged X-ray exposure in an ultrahigh vacuum could break Pb-Br bonds,[42] which may account for the fluctuations of peak intensity ratios observed in our GI-XRD results, as shown in Figure 7. In addition, cathodoluminescence (CL) intensity changes in the $MAPbBr_3$ single crystal decrease as the e-beam energy increases,[36] which aligns with the observed trend in the intensity ratio between $PbBr_2$ (111) peak and $MAPbBr_3$ (100) peak in our GI-XRD results, as shown in Figure S6. Since CL intensity is closely correlated with defect formation, and the emergence of a new $PbBr_2$ phase in our GI-XRD results is attributed to $MA^+$ vacancy formation, this suggests that, despite the fundamental differences between e-beam and X-ray beam, they could both induce similar effects on the decomposition process of the $MAPbBr_3$ single crystal.

**TABLE 2** Summary of the instability of OIHP single crystals under e-beam and X-ray beam radiation.

| Beam type | Beam condition | Crystals | Results | Refs. |
|---|---|---|---|---|
| e-beam | 5 MeV in accelerator | $MAPbBr_3$ | Pinning of the Fermi level in Kelvin probe force microscopy (KPFM) | [35] |
| | 300 keV in TEM | $MAPbI_3$ | Bubble-like flaws from TEM observation | [37] |
| | 80 and 300 keV in HRTEM and STEM | $MAPbBr_3$ and $MAPbI_3$ | Decomposition in selected area electron diffraction (SAED) and fast Fourier transform (FFT) patterns | [38–40] |
| | 2.5–30 keV in SEM | $MAPbBr_3$ | Shift in emission bands in cathodoluminescence (CL) spectra | [36] |
| | 10–15 keV in SEM | $FAPbBr_3$ and $MAPbBr_3$ | Cracking and decreased contents of C and N elements in EDS | This work |
| X-ray beam | 10 kV in X-ray gun and 150 kV in X-ray tube | $MAPbBr_3$ | Evolutions of elemental ratio in XPS | [42, 43] |
| | 100 and 630 eV in synchrotron | $FAPbBr_3$ and $MAPbBr_3$ | Peak shift and intensity decrease in in-situ XPS | [44, 45] |
| | 11–17 keV in synchrotron | $FAPbBr_3$ and $MAPbBr_3$ | Increased peak intensity of $PbBr_2$ in GI-XRD patterns | This work |

Self-healing in OIHP polycrystalline thin films has been reported after proton or gamma-ray radiation, as well as mechanical bending.[15,18,22,25] Ion migration, defect recombination, and grain-boundary transport can neutralize radiation-induced traps and enable partial recovery of photovoltaic performance.[18,22,25] Mechanical or thermal treatment can also assist crack closure and surface healing when the cracks are introduced by bending.[15] However, in our work, beam radiation induces substantial volatilization of organic components and creates high vacancy concentrations in $FAPbBr_3$ and $MAPbBr_3$ single crystals, generating significant volume strain that drives structural evolution and extensive crack formation. The massive loss of organic species results in irreversible decomposition and prevents the lattice from reforming across the fracture surfaces. Therefore, the beam radiation-induced degradation in OIHP single crystals makes the crack healing highly unlikely in our work.

## 5 | CONCLUSIONS

In this study, we investigate e-beam and X-ray beam radiation-induced instabilities of two typical OIHP single crystals, $FAPbBr_3$ and $MAPbBr_3$, using in-situ SEM imaging, EDS analysis, and synchrotron X-ray diffraction. These in-situ characterizations provide direct and real-time experimental evidence of radiation-induced cracking and composition evolutions in both OIHP single crystals, revealing distinct and material-specific degradation pathways under beam radiation. Under the e-beam radiation, we uncover contrasting cracking modes that reflect their different stability and volatilization behavior of organic cations. There are 3-point star-style cracks in $FAPbBr_3$, while bricklayer-style cracks in $MAPbBr_3$. Under the X-ray beam radiation, the in-situ GI-XRD results capture the progressive phase evolution and $PbBr_2$ formation, providing direct structural confirmation of decomposition. By exploring the effect of beam radiation on mechanical performance, we offer deep insights into the stability issues and viability concerns of OIHP single crystals. We propose a coherent mechanism from the perspective of energy conversion during degradation. The volatilization of organic components in OIHP single crystals leads to volume strain in an unstable intermediate structure, and the energy released from volume strain is converted into the energy for crack nucleation and growth. We further provide an in-depth discussion of the underlying mechanism for the distinct crack patterns and decomposition processes between $FAPbBr_3$ and $MAPbBr_3$ single crystals based on their intrinsic mechanical properties, as well as the structure and stability variations of the organic cations. Finally, we compare our findings with the literature to underscore the instability of OIHP single crystals under high-energy radiation and highlight the importance of developing radiation-resistant strategies for their practical applications.

## ACKNOWLEDGMENTS

R.C. and Y.Z. acknowledge the financial support from the Discovery Grants Program of the Natural Sciences and Engineering Research Council of Canada (NSERC) RGPIN-2018-05731, the Ontario Early Researcher Award, and the Canada Foundation for Innovation (CFI)—Evans Leaders Fund (JELF) Number 38044. Part of the research described in this paper was performed at the Canadian Light Source (CLS), a national research facility of the University of Saskatchewan, and R.C. acknowledges the receipt of support from the CLSI Student Travel Support Program. The authors thank Sal Boccia for the technical support on SEM at the Open Centre for the Characterization of Advanced Materials (OCCAM), and Nuo Qu for the help on XRD analysis. The authors also appreciate Kai Huang and Mingqiang Li for constructive discussions.

## CONFLICT OF INTEREST STATEMENT

The authors declare no competing interests.

## DATA AVAILABILITY STATEMENT

Data will be made available upon request.

## SUPPORTING INFORMATION

Additional supporting information can be found online in the Supporting Information section at the end of this article.